\documentclass[copyright]{eptcs}
\providecommand{\event}{T. Neary, D. Woods, A.K. Seda and N. Murphy (Eds.):\\ The Complexity of Simple Programs 2008.}
\providecommand{\volume}{1}
\providecommand{\anno}{2009}
\providecommand{\firstpage}{226}
\usepackage{breakurl}
\usepackage{amsmath}
\usepackage{amssymb}
\usepackage{stmaryrd}
\usepackage{url}
\usepackage{graphicx}
\usepackage{xspace}
\usepackage{theorem}

\theoremstyle{plain}
\theorembodyfont{\slshape}
\newtheorem{theorem}{Theorem}[section]

\newtheorem{proposition}{Proposition}[section]

\newcommand{\Limp}{\,\operatorname{\Rightarrow}\,}

\graphicspath{{./}}

\def\observe#1{\mathsf{O}_{#1}}
\def\nstep{\shortrightarrow}

\def\class#1{\mathbb{C}_{#1}}
\def\cC{\mathcal{C}}

\def\cR{\mathcal{R}}
\def\cP{\mathcal{P}}
\def\cN{\mathcal{N}}
\def\Z{\mathbb{Z}}
\def\N{\mathbb{N}}
\def\L{\mathcal{L}}            

\def\mathcorr{\ifx\next.\mkern -.3mu\else\ifx\next,\mkern -.3mu\else\,\fi\fi}
\def\func#1#2#3{{#1}:{#2}\to{#3}\futurelet\next\mathcorr}
\def\pair#1#2{ \langle #1,#2 \rangle\futurelet\next\mathcorr }
\providecommand{\Star}[1]{{#1}^{\star}}

\def\mathbold#1{\expandafter\def\csname#1\endcsname{\mathbf{#1}}}
\mathbold{2}

\def\set#1#2{\ensuremath{\{\, {#1} \mid {#2}\,\}}}
\def\setof#1{\left\{ {#1} \right\}}
\def\card#1{\left|{#1}\right|} 

\def\redT{\leq_T}
\def\redsT{<_T}

\def\Kleene#1{\setof{#1}}
\def\re{\relax\ifmmode \mathrm{r.e.} \else r\@.e\@.\xspace\fi}
\def\ce{c\@.e\@.\xspace}

\title{Computational Processes and Incompleteness}

\author{Klaus Sutner
\institute{Carnegie Mellon University\\Pittsburgh PA 15213, USA}
\email{\url{http://www.cs.cmu.edu/~sutner}}
}
\def\titlerunning{Computational Processes and Incompleteness}
\def\authorrunning{K. Sutner}

\begin{document}
\maketitle

\begin{abstract}
We introduce a formal definition of Wolfram's notion of computational 
process based on cellular automata, a physics-like model of computation.  
There is a natural 
classification of these processes into decidable, intermediate and complete. 
It is shown that in the context of standard finite injury priority arguments 
one cannot establish the existence of an intermediate computational process. 
\end{abstract}


\section{Computational Processes}
\label{sec:comp-proc}

Degrees of unsolvability were introduced in two important papers by 
Post \cite{Post44:re_sets_decision_problems}
and Kleene and Post \cite{KleenePost54:upper_semi_lattice_degrees}.
The object of these papers was the study of the complexity of decision 
problems and in particular their relative complexity: how does a 
solution to one problem contribute to the solution of another, a 
notion that can be formalized in terms of Turing reducibility and 
Turing degrees.  
Post was particularly interested in the degrees of recursively enumerable 
(\re) degrees.  
The Turing degrees of \re sets together with Turing reducibility form 
a partial order and in fact an upper semi-lattice $\cR$.
It is easy to see that $\cR$ has least element 
$\mathbf{\emptyset}$, the degree of decidable sets, and a largest element 
$\mathbf{\emptyset}'$, the degree of the halting set. 
Post asked whether there are any other \re degrees and embarked on a 
program to establish the existence of such an  \emph{intermediate degree} by 
constructing a suitable \re set.  
Post's efforts produced a number of interesting ideas such as simple, 
hypersimple and hyperhypersimple sets but failed to produce an intermediate degree.
 
The solutions to Post's problem required a new technique now called a 
\emph{priority argument} that has since become one of the hallmarks of 
computability theory.  
Interestingly, the technique was invented 
independently and nearly simultaneously by Friedberg and Muchnik, 
see \cite{Friedberg57:incomparable,Muchnik56:incomparable}. 
Our structural understanding of $\cR$ 
has grown significantly over the last half century, see 
\cite{Soare87:book} for a slightly dated but excellent overview or 
\cite{AmbosSpiesFejer06:degrees} for a more recent account. 
For example, it is known that every countable partial order can be embedded 
into $\cR$ which fact leads to the decidability of the $\Sigma_1$ theory 
of $\cR$. 
According to a theorem by Sacks, the partial order of the \re 
Turing degrees is dense: whenever $A \redsT B$ for two \re sets 
$A$ and $B$ there is a third \re set such that $A \redsT C \redsT B$,
\cite{Sacks64:density}.
Overall, the first order theory of $\cR$ is highly undecidable 
\cite{HarringtonShelah82}.

More recently, in his book \cite{Wolfram02:anks}, Wolfram suggests that 
``\ldots all processes, whether they are produced by human effort or 
occur spontaneously in nature, can be viewed as computations.''
This assertion is not particularly controversial, though it does require 
a somewhat relaxed view of what exactly constitutes a computation--%
as opposed to an arbitrary physical process such as, say, a waterfall. 
Wolfram then goes on to make a rather radical proposition, the so-called 
\emph{Principle of Computational Equivalence} (PCE, for short): 
``\ldots almost all processes that are not obviously simple can be viewed 
as computations of equivalent sophistication.''
The reference contains no definition of what exactly constitutes a 
computational process, or what is meant by sophistication, 
so it is a bit difficult to take issue with the assertion.  
However, most recursion theorists would agree that PCE coexists uneasily
with the well-established theory of the \re degrees.
Wolfram's response to such criticism can be summarized thus \cite{Wolfram02:priv}: 
any of the standard constructions of an intermediate \re set achieves 
precisely that, it constructs a particular \re set that is undecidable 
yet fails to be complete.  
However, the construction as a whole, when interpreted 
as a computational process, is very different from the set so constructed. 
Most notably, the process makes heavy use of universal Turing machines 
and may thus well be complete when viewed in its entirety.
For example, the standard Friedberg-Muchnik construction produces two 
\re sets $A$ and $B$ that are incomparable with respect to Turing reductions.
Hence, each individual set so constructed has intermediate degree. 
However, it was shown by Soare that the disjoint sum $A \oplus B$ is 
in fact complete, see \cite{Soare72:fried_muchn}. 
It is thus hard to see how this priority argument could be construed as not 
being a complete process.

The difficulty in identifying intermediate processes is closely related
to another often observed problem with intermediate degrees:
all existence proofs of 
intermediate degrees are artificial in the sense that the constructed \re 
sets are entirely ad hoc.  
This was perhaps stated most clearly by M. Davis \cite{Davis03:natural_degrees}: 
``But one can be quite precise in stating that no one has 
produced an intermediate \re degree about which it can be said that 
it is the degree of a decision problem that had been previously 
studied and named.''
More recently, Ambos-Spies and Fejer in \cite{AmbosSpiesFejer06:degrees} 
are equally forceful about the lack of natural intermediate problems: 
``The sets constructed by the priority method to solve Post's Problem 
have as their only purpose to be a solution. \ldots Thus it can be said 
that the great complexity in the structure of the \ce degrees arises 
solely from studying unnatural problem.'' 
Note that results from degree theory have been transfered to other areas. 
For example, Boone \cite{Boone65:groups_degrees} has shown how to construct 
a finitely presented group whose word problem has a given, arbitrary \re degree. 
Of course, the translation itself is entirely 
unsuspicious, it is only the instantiation required in particular to 
produce intermediate degrees that is unsatisfactory. 
In a similar vein, it was shown by Feferman that derivability in a formal theory fully 
reflects the structure of the \re degrees: for every \re degree $\mathbf{d}$ 
there is an axiomatizable theory whose collection of theorems has 
degree exactly  $\mathbf{d}$, see \cite{Feferman57:degrees_unsolvability}.
For a more recent and more complicated application of degree theory to 
differential geometry see Soare's contribution to the proof of Gromov's 
theorem, \cite{Soare04:diff_geometry}.

There appears to be a fairly strong connection between the lack 
of natural intermediate degrees and PCE.  Indeed, any of Davis' 
previously studied and named decision problems could presumably be 
translated into an intermediate computational process. For example, 
following a suggestion by H. Friedman, suppose one can identify a 
simple formula $\varphi$ of, say, Peano arithmetic such that 
$\mathrm{Th}(\varphi)$ has intermediate degree.  
Any standard enumeration of all theorems provable from $\varphi$ 
would then constitute an intermediate computational process.  
Of course, at present no one knows how to construct such a formula.

The purpose of this paper is to introduce a plausible definition 
of a computational process
and to study the existence of intermediate processes in this framework. 
In section \ref{sec:comp-proc-cell-aut} we give a definition 
that is modeled closely on discrete dynamics and in particular 
on cellular automata. We will show that some complexity results 
transfer over to the computational process point of view, while 
others do not. 
In section \ref{sec:constr-interm-sets} we consider directly finite injury 
priority arguments, the traditional source of intermediate \re sets. 
As we will see, at least the standard minimal construction fails to produce 
an intermediate process. 
We discuss possible further directions and open problems in the last 
section.  
To keep this paper reasonably short we will refrain from introducing 
standard concepts from recursion theory and refer the reader to 
texts such as \cite{Rogers67:book,Soare87:book,Cooper04:book}.

\section{Computational Processes and Cellular Automata}
\label{sec:comp-proc-cell-aut}

It stands to reason that any definition of a computational process in 
Wolfram's sense should capture some aspect of a physical computation 
rather than the purely logical structure of such a computation.
Note that feasibility is not an issue here, though: we are not 
concerned with computations that could be accomplished within certain 
space, time or energy bounds such as the ones that prevail in our 
physical universe. 
Rather, we assume an unlimited supply of memory and unbounded time, 
as is customary in the abstract theory of computation. 
When it comes to selecting a physics-like model of computation 
the examples in \cite{Wolfram02:anks}, as well as previous work 
by Wolfram, suggest cellular automata, see also \cite{Ilachinski01:cellular_automata}.
Technically these are continuous shift-invariant maps on Cantor spaces 
of the form  $\Sigma^{\Z^{d}}$ where $\Sigma$ is some finite alphabet. 
For simplicity we will only consider the one-dimensional case $d=1$ here. 
Thus we have a \emph{global map} $G$ acting on $\Sigma^{\Z}$, the 
space of all \emph{configurations}. 
By continuity and shift-invariance $G$ is induced by a finite map 
$\func{g}{\Sigma^{w}}{\Sigma}$ where $w \geq 1$, the so-called local map of 
the automaton.  
Thus, a cellular automaton has a finite description. 
Note that $G$ can be computed by a rational transducer operating 
on bi-infinite words. 
To avoid issues with higher-type computability we focus on the subspace 
$\Sigma^{\Z}_{0}$  of configurations $X$ of finite support: 
suppose $0 \in \Sigma$ and allow only finitely many 
positions $i \in \Z$ such that $X(i) \neq 0$.
We may safely assume that the global map preserves finite supports. 
Hence we can identify configurations with finite words over $\Sigma$ 
and the orbit of any such configuration is automatically \re

It is quite straightforward to simulate Turing machines in this setting, 
yet in a sufficiently constrained framework such as first order logic 
with equality no undecidability issues arise. 
More precisely, for a one-dimensional cellular automaton $g$ 
consider the structure 
$\cC_{g} = \pair{\Sigma^{\Z}_{0}}{\nstep}$
where $\nstep$ is the binary relation representing a single application 
of the global map.  This representation is preferable for purely technical 
reasons; notably, there are only finitely many atomic formulae over 
a finite set of variables in this setting. 

\begin{theorem}
\label{thm:first-order}
For one-dimensional cellular automata, first order logic $\L(\nstep)$ 
with equality is decidable.
\end{theorem}

We note in passing that this result also holds over the full 
space $\Sigma^{\Z}$ of infinite configurations, though the 
machinery required to establish decidability is quite a bit more 
complicated, see \cite{Sutner07:model_check}.
Dimensionality is crucial here, no similar result holds in dimensions 
two or higher.

\begin{figure}[h]
  \begin{center}
    \includegraphics[scale=10.0]{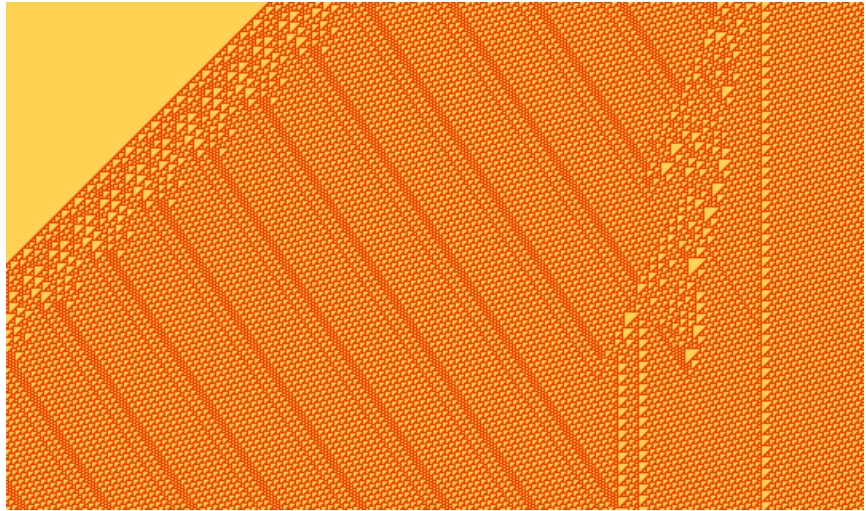}
  \end{center}
  
  \caption{A partial orbit of elementary cellular automaton number 110.}
\label{fig:eca110}
\end{figure}

At any rate, we can now propose a formal definition of a computational 
process. 
For our purposes, a \emph{computational process} over some 
alphabet $\Sigma$ is a 
pair $P = \pair{\tau}{X}$ where $\tau$ is a finite state transducer over 
$\Sigma$ and $X \in \Star{\Sigma}$ a word. 
We will conflate $\tau$  and the map that it induces on $\Star{\Sigma}$
and refer to $\tau$ as the \emph{computor} and to $X$ as 
the initial configuration of $P$. 
Thus we can obtain a sequence of configurations by iterating 
$\tau$ on $X$: $X_{t} = \tau^{t}(X)$. 
As Wolfram's work in the early 1980's showed, a careful study of the 
images obtained by plotting part of an orbit $(X_{t})_{t \geq 0}$
as a two-dimensional grid can produce some degree of insight in 
the computational properties of the system. 
For example, figure \ref{fig:eca110} shows part of an orbit of elementary 
cellular automaton (where the local map is of the form $\2^3 \to \2$)
number 110.

The geometric structure is rather surprisingly complicated and there 
are lots of persistent, localized structures. 
It is not inconceivable that, given proper initial conditions, this type 
of behavior might be exploited to perform computations. 
This turns out to be indeed the case as shown by Cook
\cite{Cook04:universal_110}; the proof is quite difficult, however. 
In fact, it would be rather challenging to present this argument in 
a purely non-geometric way suitable by verification through a proof-checker.

In order to model the extraction of information from an orbit we define 
an \emph{observer} to be a word function $\rho$ on $\Sigma$ that is 
computable in constant space.  
We associate the 
language $\observe{\rho} = \bigcup \set{ X_{t}\rho}{ t \geq 0 }$ 
with the observer.
We refer to $\observe{\rho}$ as the \emph{observation language} of $\rho$. 
It follows from the definitions that this language is \re

The computor is narrowly constrained so as to make sure that a 
single step in the process is very simple and in particular cannot 
hide complicated sub-computations. 
For example, the computor cannot perform a whole stage in a Friedberg-Muchnik
type construction in a single step, see below. 
The observer on the other hand has the ability to filter out some part 
of the current configuration and rewrite it slightly. 
It is important that the observer is strictly constrained in computational 
power.  For example, suppose we were to allow an arbitrary primitive recursive 
word function to be applied to $X_{t}$.  
The observer could then ignore the input and simply launch an independent 
computation of his own. 
In particular, as long as $\limsup \card{X_{t}} = \infty$ the observer could 
always produce a complete observation language.  
Thus even an entirely trivial computational process where 
$X_{t} = 0^{t}$ would admit a maximally complicated observation. 

We can now classify computational processes according to the complexity 
of the associated observations. 
A computational process is \emph{undecidable} if there exists an observer 
whose observation language is undecidable. 
Likewise, 
a computational process is \emph{complete} if there exists an observer 
whose observation language is \re-complete. 
The process is \emph{intermediate} if it is undecidable but 
fails to be complete. 
Thus, an intermediate computational process admits at least one observer 
that finds the process undecidable but prohibits any observer from extracting 
a complete \re set.


As theorem \ref{thm:first-order} suggests, undecidability requires us 
to consider whole orbits rather than just fixed size segments. 
In any system such as monadic second order logic or transitive closure logic 
that allows us to express the Reachability Problem for $\cC_{\rho}$:  
``Does configuration $y$ appear in the orbit of configuration $x$?'' 
we should expect undecidable propositions. 
Note that since we are dealing with configurations of finite support 
the Reachability Problem is \re 
Also, if we think of the cellular automaton as a computational process
starting at $x$ there is an observer that can check whether $y$ appears.

Since our notion of computational process is motivated by cellular automata 
it is worthwhile to consider standard classifications of these systems 
and how they relate to computational processes. 
Classifications of cellular automata are more
or less based on attempts to transport concepts from classical dynamics into 
the realm of cellular automata, perhaps augmented by ideas from recursion 
theory, see 
\cite{Wolfram84:computation_theory,Wolfram02:anks,LiPackard90:structure_elementary_ca,LiPackardLangton90:transition,Kurka97:equicontinuity,Kurka03:topol_symb_dynamics}. 
As one might expect, from a computational perspective these are all 
fraught with undecidability. 
For example, it is $\Pi^0_2$-complete to determine whether all orbits 
end in a fixed point, 
it is $\Sigma^0_3$-complete to determine whether all orbits 
are decidable, and it is $\Sigma^0_4$-complete to determine whether 
a given cellular automaton is computationally universal, 
see \cite{Sutner89:Culik_Yu,Sutner90:classca}. 
It was suggested in \cite{Sutner03:intermediate} to turn this difficulty 
into a tool: one can use the complexity of Reachability to classify cellular 
automaton in a much more fine-grained manner than usual.

To this end, define $\class{\mathbf{d}}$ to be the collection of 
all cellular automata whose Reachability Problem has degree 
exactly $\mathbf{d}$, some \re degree. 
Note that this is a significantly stronger condition than 
having a cellular automaton that is capable of, say, enumerating 
a set whose degree is $\mathbf{d}$:  only a few of the orbits of 
the cellular automaton will correspond to actual computations of 
the corresponding Turing machine (or whatever other computational 
model the automaton may emulate).  It requires a bit of care to 
make sure that the other orbits do not violate the degree condition. 
At any rate, we have the following result, 
see \cite{Sutner02:intermediate,Sutner03:intermediate}.

\begin{theorem}{Degree Theorem \\}
For every \re degree $\mathbf{d}$ there is a one-dimensional 
cellular automaton whose Reachability Problem has degree precisely $\mathbf{d}$.  
In fact, the cellular automaton can be chosen to be reversible. 
\end{theorem}

Thus, the whole complexity of the upper semi-lattice $\cR$ of the 
\re degrees is fully reflected in this classification of cellular automata. 
In fact, the situation is even a bit more complicated: we can fine-tune 
various computational aspects of the cellular automaton in question. 
For example, consider the natural Confluence Problem: 
``Given two configurations, do their orbits overlap?''.  

\begin{theorem}{Two Degree Theorem \\}
For any \re degrees $\mathbf{d}_{1}$ and $\mathbf{d}_{2}$, there is a 
one-dimensional cellular automaton whose Reachability Problem has degree 
$\mathbf{d}_{1}$ and whose Confluence Problem has degree $\mathbf{d}_{2}$.
\end{theorem}

Of course, the Two Degree theorem cannot possibly hold when one considers 
only reversible cellular automata: in this case confluence of $x$ and $y$ 
is equivalent with $x$ being reachable from $y$ or $y$ being reachable from $x$.

Needless to say, the proofs given in the references for all these degree related 
results on cellular automata are based on the existence of intermediate \re sets. 
More precisely, they employ somewhat complicated simulations of an 
arbitrary Turing machine $M$ by a cellular automaton that ensures that the degree 
of the Reachability problem for the cellular automaton is precisely the 
same as the degree of the acceptance language of $M$. 
If we were to to reinterpret these arguments in terms of computational processes 
the process would have to include the simulation of a Turing machine with 
intermediate acceptance language. 
It is these sub-computations that push the degree of the Reachability problem 
up to the chosen degree, and it is the overall construction using restarting 
Turing machines that keep it from exceeding the target degree. 
Alas, it is not hard to construct an observer that extracts details of the 
computation performed by $M$.  As we will see in the next section, at least 
for the standard choice of $M$, this means that the computational process
admits a complete observation and is thus not intermediate.

\section{Constructing Intermediate Sets}
\label{sec:constr-interm-sets}

In order to spoil a potential observer that is trying to extract additional information
from a process it seems reasonable to rely on 
the least complicated construction of an intermediate degree known.   
In this section we will discuss what appears to be the most basic construction. 
Our description here will be rather terse; we refer the reader to 
\cite{Soare87:book,Odifreddi99:recursion_theory_ii} for background. 

As already mentioned, the classical Friedberg-Muchnik construction establishes the 
existence of two Turing incomparable \re sets. 
Here we use a method that is slightly less ambitious. 
It produces two sets $S$  and  $A$  such that 
\begin{align*}
   \cP_{e} &:\qquad  W_{e} \text{ infinite } \Limp  S \cap W_{e} \neq \emptyset \\
   \cN_{e} &:\qquad  A \not\simeq  \Kleene{e}^{S}
\end{align*}
for all $e \in \N$.  
As is customary, we have identified a set with its characteristic function. 
Here $S$ is the set we are trying to construct and $A$ is an auxiliary \re set 
that demonstrates that $S$ is incomplete. 
The two sets are constructed in stages $s < \omega$ and the actions taken 
at each stage are easily seen to be primitive recursive in $s$, 
so the sets are indeed \re

The stated conditions are referred to as \emph{requirements}.  
More precisely, $\cP_{e}$ is a positive requirement that can be satisfied by 
placing elements into $S$. 
The negative requirements $\cN_{e}$ are a bit more complicated: we have to make 
sure that the $e$th partial recursive function using $S$ as an oracle cannot 
be the characteristic function of $A$. 
Hence, the positive requirements insure that $S$ is simple and 
thus undecidable whereas
the negative requirements guarantee that $A \not\redT S$ so that $S$ 
cannot be complete. 
The principal problem with this construction is that the requirements may well 
conflict with each other. 
E.g., we may place an element $x$ into $S$ to satisfy $\cP_{e}$. 
But then the oracle in $\Kleene{e'}^{S}$ changes and it may happen that 
$A(z) \simeq  \Kleene{e}^{S}(z)$ where previously there was a disagreement. 

In order to deal with this problem one has to exercise a high degree of control 
over the way elements are enumerated into $S$ and $A$. 
The key idea is to order all requirements into an $\omega$-sequence
$$
     \cP_{0} <  \cN_{0} < \cP_{1} <  \cN_{1} < \cP_{2} <  \ldots 
$$
where lower rank means higher priority.
At any stage during the construction, we work only on the requirement of 
highest priority that currently fails to be satisfied and that can be 
addressed at this stage. 

The positive requirements are relatively easy to deal with: $\cP_{e}$ requires 
attention at stage $s \geq e$ if  $S^{<s} \cap W^{<s}_{e} = \emptyset$ and  
there is some $x \in W^{s}_{e}$ such that $x > 2e$ and $x$ does not violate any 
constraints imposed by requirements of higher priority.   
If we act on this requirements, we place the least such $x$  into $S$. 

The strategy to deal with negative requirements is to allocate an unbounded set of 
unique potential witnesses for each requirement $\cN_{e}$ with the intent of possibly 
placing a witness into $A$.  
For the sake of definiteness choose $\set{\pair{e}{u}}{u \in \N}$ 
where $\pair{a}{b}$ is a standard pairing function. 
The requirement is satisfied at stage $s$ when, for the current witness 
$w_{e} = \pair{e}{u}$, we have 
$$
        A^{<s}(w_{e}) \not\simeq \Kleene{e}^{S^{<s}}_{s}(w_{e})
$$
Otherwise the requirement requires attention.
In this case we choose a new witness $x$
that is larger than any number so far used in the construction and 
we try to keep $x$ out of $A$ (i.e., we do nothing). 
If, at some later stage $t$, we find that 
$$
        \Kleene{e}^{S^{<t}}_{t}(w_{e}) \simeq 0
$$
we place $w_{e}$ into $A$ and protect this computation in the sense 
that we associate with $\cN_{e}$ as constraint the largest number 
used in the computation of $\Kleene{e}^{S^{<t}}_{t}(w_{e})$.

Note that action taken on behalf of some requirement may well cause 
other requirements of lower priority to become unsatisfied. 
Since, for each requirement, there are only finitely many of higher priority 
the requirement will not be injured any more after an initial segment of 
the construction; this is easy to show by induction. 
It follows that the construction is successful and $S$ has intermediate degree, 
as intended. 
Unfortunately, as in the classical Friedberg-Muchnik construction, 
we are computing more that just $S$, in particular the construction 
also yields the auxiliary set $A$. 

\begin{proposition}
The disjoint union $S \oplus A$ is complete.
\end{proposition}

To see this note that the collection of witnesses can be viewed as an $S_{2}$-function 
$w(e,s)$ such that the limit $w(e) = \lim w(e,s)$ exists for all $e$. 
But $w(e)$ is either the largest element in $\pair{e}{\N} \cap A$ or the least 
element in $\pair{e}{\N} - A$.  Using $S$ as an oracle one can decide which case 
holds, so that $e \mapsto w(e)$ is recursive in $S \oplus A$. 
Now choose a recursive function $f$ such that 
$$
    e \in S'  \iff  \Kleene{f(e)}^{S}(z) \simeq 0  \text{  for all }z
$$
But then $e \in S' \iff  \Kleene{f(e)}^{S}(w(e)) \simeq 0 \iff w(e) \in A$
so that $S' \redT S \oplus A$. 
\vspace{1ex} 

Similar arguments seem to apply to all priority constructions.  
In fact, it was suggested by Jockusch and Soare in 
\cite{JockuschSoare72:degrees_pi_class}
that priority constructions obey a kind of ``maximum degree principle'' in 
the sense that the construction of an \re set $S$  with weak negative requirements
automatically produces a complete set. 
If the requirements are strong enough to prevent completeness of $S$ there 
are still remnants of completeness in the construction.
This type of completeness is hidden in the standard recursion theoretic argument
but becomes visible when we recast the argument as a computational process. 

In the case of a priority argument this translation is straightforward in principle 
but rather too tedious to be carried out in reality in any detailed manner 
(for example, by explicitly constructing the transducer for the computor). 
One possible approach is to first rephrase the construction as a Turing 
machine and then to code the instantaneous descriptions of the Turing machine 
as words over some suitable alphabet.  
Note that in this setting a single step in the computation can indeed be 
computed by a transducer. 
There are several ways to deal with the intended output $S$. 
For example, one could mimic a write-only output tape by using multiple tracks. 
The appropriate observer $\rho$ can then simply project from this virtual output tape.  
The observation language $\observe{\rho}$ is clearly Turing equivalent to $S$. 
Alas, from the perspective of the computational process there is little 
difference between $S$ and $A$: both are sets of natural numbers 
and finite monotonic approximations to these sets are generated as the 
process unfolds according to the rules of the priority argument. 
Thus, a constant space observer is powerful enough to produce an 
observation language that is Turing equivalent to $S \oplus A$. 
Thus we have the following result.

\begin{theorem}
The simplified Friedberg-Muchnik priority argument does not 
yield an intermediate computational process. 
\end{theorem}

There are  variants of this construction that produce stronger results. 
For example, Sacks \cite{Sacks63:degrees_less_emptyjump}
has shown how to modify the construction so that $A$ can be any given 
undecidable \re set and we obtain $S$ such that  $A \not\redT S$. 
Thus $S$ avoids a whole upper cone rather than just complete sets. 
Alas, as a computational process the construction requires us to furnish 
an enumeration of $A$ which apparently requires a complete process.

\section{Open Problems}
\label{sec:open-problems}

We have proposed a notion of computational process that 
is motivated by discrete dynamical systems that provide a physics-like 
model of computation.  
We have verified that standard finite injury priority arguments, when 
interpreted as computational processes in this sense, produce 
complete rather than intermediate processes. 
In light of the Jockusch/Soare maximum degree principle it is difficult 
to imagine a variant of a priority argument that would circumvent this problem. 
We conjecture that this obstruction is truly general.

A number of questions come to mind.  
First, one could try to use a general framework for priority arguments
such as the Baire categoricity approach and try to establish that 
all such attempts at constructing intermediate processes fail. 
The problem here is that though the framework accommodates a variety of 
priority constructions it is far from clear which class of constructions 
yield intermediate sets and are thus suitable candidates. 
Second, it would be interesting to constrain the model of computation 
that motivates our definition of computation process further in the direction 
of making it a closer match to actual physical computation. 
For example, one could consider higher-dimensional cellular automata 
that are reversible and in addition obey certain conservation laws, 
much in the spirit of Fredkin's recent work on SALT 
\cite{FredkingMiller05:salt}.  
The motivation for Fredkin is the construction of physically feasible, 
three-dimensional systems that dissipate as little energy as possible.
It is conceivable that a narrow class of such systems could maintain 
universality while eliminating intermediate processes. 
Of course, the existence of a ``natural'' intermediate degree would 
derail such efforts. 
It should be noted that the first order theory of two-dimensional cellular automata 
is already undecidable, so some additional care may have to be taken to make sure that 
the one-step relation in a computational processes suitable to model such devices 
is sufficiently simple.  
Third, Kucera gave a priority-free solution to Post's Problem, 
see \cite{Kucera86:priority_free}.
Their argument relies on the low basis theorem due to Jockusch and Soare 
\cite{JockuschSoare72:pi_zero_one_classes} whose proof appears to require 
universality.  Moreover, it is not entirely clear how to 
rephrase the entirety of Kucera's argument as a computational process.  
Lastly, one may question whether our definition of computational process 
is indeed appropriate.  Any attempt to strengthen the computor seems to permit 
the existence of intermediate processes, albeit for the wrong reasons: 
the computor can hide essential parts of the computation from the observer. 
Strengthening the observer is similarly difficult since it tends to 
produce only complete processes, again for the wrong reasons.

\bibliographystyle{eptcs}



\end{document}